\documentclass[
reprint,
superscriptaddress,
preprint,
onecolumn,
showpacs,preprintnumbers,
amsmath,amssymb,
aps,
acknowledgement,
]{revtex4-1}

\usepackage{dcolumn}
\usepackage{bm}
\usepackage{color}
\usepackage[dvipdfmx]{graphicx}
\usepackage{amsmath}
\usepackage{ulem}
\usepackage{fancyhdr}
\usepackage{titlesec}

\newcommand{\FMO}{Fe$_{2}$Mo$_{3}$O$_{8}$ }

\newcommand{\FZMO}{(Fe$_{0.95}$Zn$_{0.05}$)$_{2}$Mo$_{3}$O$_{8}$ }
\newcommand{\dSs}{$\Delta S^{*}$ }
\newcommand{\dMs}{$\Delta M^{*}$ }
\newcommand{\Hs}{$H^{*}$ }

\renewcommand{\figurename}{{\bf Fig.}}

\makeatletter
\def\fnum@figure{\figurename~{\bf \thefigure} }
\def\@caption@fignum@sep{\textbar \space}
\makeatother

\titleformat*{\section}{\large\bfseries}
\titleformat*{\subsection}{\normalsize\bfseries}
\titleformat*{\subsubsection}{\normalsize\bfseries}

\graphicspath{{./}}

\begin{document}
\setcitestyle{super}
\makeatletter 
\def\@biblabel#1{#1. }
\makeatother

	\title{Thermodynamic determination of the equilibrium first-order phase-transition line hidden by hysteresis in a phase diagram}
	\author{Keisuke Matsuura}
	\email[E-mail:]{keisuke.matsuura@riken.jp}
	\affiliation{RIKEN Center for Emergent Matter Science, Wako 351-0198, Japan}

	\author{Yo Nishizawa}
	\affiliation{Department of Applied Physics and Quantum-Phase Electronics Center (QPEC), University of Tokyo, Tokyo 113-8656, Japan}
	
	\author{Markus Kriener}
	\affiliation{RIKEN Center for Emergent Matter Science, Wako 351-0198, Japan}
	
	\author{Takashi Kurumaji}
	\affiliation{Department of Advanced Materials Science, University of Tokyo, Kashiwa 277-8561, Japan}
	
	\author{Hiroshi Oike}
	\affiliation{RIKEN Center for Emergent Matter Science, Wako 351-0198, Japan}
	\affiliation{Department of Applied Physics and Quantum-Phase Electronics Center (QPEC), University of Tokyo, Tokyo 113-8656, Japan}
	\affiliation{PRESTO, Japan Science and Technology Agency (JST), Kawaguchi 332-0012, Japan}
	
	\author{Yoshinori Tokura}
	\affiliation{RIKEN Center for Emergent Matter Science, Wako 351-0198, Japan}
	\affiliation{Department of Applied Physics and Quantum-Phase Electronics Center (QPEC), University of Tokyo, Tokyo 113-8656, Japan}
	\affiliation{Tokyo College, University of Tokyo, Tokyo 113-8656, Japan}
	
	\author{Fumitaka Kagawa}
	\email[E-mail:]{kagawa@phys.titech.ac.jp}
	\affiliation{RIKEN Center for Emergent Matter Science, Wako 351-0198, Japan}
	\affiliation{Department of Applied Physics and Quantum-Phase Electronics Center (QPEC), University of Tokyo, Tokyo 113-8656, Japan}
	\affiliation{Department of Physics, Tokyo Institute of Technology, Tokyo 152-8551, Japan}
	
	\date{\today}
	
\begin{abstract}
	Phase diagrams form the basis for the study of material science, and the profiles of phase-transition lines separating different thermodynamic phases include comprehensive information about thermodynamic quantities, such as latent heat. However, in some materials exhibiting field-induced first-order transitions (FOTs), the equilibrium phase-transition line is hidden by the hysteresis region associated with the FOT; thus, it cannot be directly determined from measurements of resistivity, magnetization, etc. Here, we demonstrate a thermodynamics-based method for determining the hidden equilibrium FOT line. This method is verified for the FOT between antiferromagnetic and ferrimagnetic states in magneto-electric compounds (Fe$_{0.95}$Zn$_{0.05}$)$_{2}$Mo$_{3}$O$_{8}$. The equilibrium FOT line determined based on the Clausius--Clapeyron equation exhibits a reasonable profile in terms of the third law of thermodynamics, and it shows marked differences from the midpoints of the hysteresis region. Our findings highlight that care should be taken for referring to the hysteresis midpoint line when discussing field-induced latent heat or magnetocaloric effects.
\end{abstract}

\maketitle

\pagestyle{fancy}
\lhead{}
\chead{}
\rhead{}
\rfoot{\thepage}
\cfoot{}
\renewcommand{\headrulewidth}{0pt}

Condensed matter often changes its structural/electric/magnetic states in response to changes in temperature, $T$, and external fields such as magnetic fields, $H$, electric fields, $E$, and pressures, $P$. These changes are concisely summarized in phase diagrams. In addition to separating different phases, the profile of a phase-transition line includes comprehensive information on thermodynamic quantities associated with the phase transition. For example, in the $P$--$T$ phase diagram of $^{3}$He, the slope of the first-order transition (FOT), $\frac{dT^{*}}{dP^{*}}$ (the asterisk represents a value on the phase boundary), between the solid and liquid phases becomes negative below 0.32 K, indicating that the solid $^{3}$He has larger entropy than the liquid $^{3}$He, which is a behaviour known as the Pomeranchuk effect \cite{Richardson1997}. Additionally, in the $P$--$T$ phase diagrams in certain organic conductors, the slope of the FOT line between antiferromagnetic (AFM) and superconducting (SC) phases is negative, whereas the slope between quantum spin liquid (QSL) and SC phases is almost perpendicular to the pressure axis. This observation indicates that the QSL phase has larger entropy than the AFM phase \cite{Pustogow2018, Furukawa2018}. Furthermore, for the case of an FOT, the Clausius--Clapeyron equation enables a quantitative estimation of the entropy change (or latent heat) accompanying the FOT, $\Delta S^{*}$ (or $T^{*}$$\Delta S^{*}$), if the change in the other extensive quantity is given (the magnetization change, $\Delta M^{*}$, for the case of an $H$--$T$ phase diagram). For instance, accurate evaluation of magnetic-field-induced latent heat has been a central issue in the study of magnetocaloric effects \cite{franco2012}, and thus, comparison with theoretical values derived from the correct phase diagram provides important insights. Thus, the accurate determination of a phase-boundary profile can be an important issue in many contexts.

However, in a real material, an FOT does not necessarily occur at the equilibrium FOT line but in a parameter region away from it. This behaviour originates from the fact that an FOT generally involves nucleation and growth, which are nonequilibrium kinetic processes. As a result, two hysteresis lines, which we define as the lower-field and higher-field boundaries of the hysteresis region, are often drawn instead of one equilibrium FOT line, particularly in an experimentally determined phase diagram. In fact, the precise determination of the equilibrium FOT line is often not straightforward in experiments.

We consider two prototypical cases of a field-temperature phase diagram that exhibits an FOT (Figs.~1{\bf a} and {\bf b}). The first case represents an FOT between ``up'' and ``down'' phases connected by a symmetry operation, such as time-reversal or space-inversion operations (Fig.~1{\bf a}). Typical ferromagnetic (FM) and ferroelectric materials are classified into this case if uniform magnetic and electric fields are chosen as the external fields, respectively. The free energies of the two states obviously degenerate in the absence of a symmetry-breaking field \cite{Chaikin2013}, and the equilibrium FOT line lies on the zero-field line by definition. The hysteresis lines in positive- and negative-field regions should be symmetric regarding the zero-field line, and thus, one may determine the equilibrium FOT points by taking the midpoints of the two hysteresis lines. In the second case, by contrast, the competing two phases do not degenerate at zero field (Fig.~1{\bf b}). A typical example includes a magnetic-field-induced FOT from AFM to FM phases. In such a case, there is no symmetry reason for the midpoints of the two hysteresis lines to agree with the equilibrium FOT points. To the best of the authors' knowledge, the validity of considering the midpoints as the FOT points has never been carefully discussed for a field-induced FOT.

The uncertainty of a field-induced equilibrium FOT line may be crucial for a material exhibiting a wide hysteresis; for instance, various materials, such as Gd$_{5}$Ge$_{4}$ \cite{Roy2007}, LaFe$_{12}$B$_{6}$ \cite{Fujieda2017}, doped CeF$_{2}$ alloys \cite{Manekar2001, Kumar2006}, doped Mn$_{2}$Sb \cite{Kushwaha2008, Singh2019}, doped manganese oxides \cite{Tokura2006, Rawat2007, Matsuura2021}, and martensitic materials \cite{Ito2008,niitsu2020} are known to exhibit distinct hysteresis broadening at low temperatures. Similar hysteresis broadening is also observed in a pressure-induced liquid-liquid transition of an aqueous solution \cite{suzuki2022}. In the present study, by focusing on (Fe$_{1-y}$Zn$_{y}$)$_{2}$Mo$_{3}$O$_{8}$ (Fig.~1{\bf c}), we exemplify how the equilibrium FOT line is determined for a material exhibiting a broad hysteresis and show that the midpoints of the two hysteresis lines appreciably deviate from the equilibrium FOT line. This finding indicates that considering the midpoints of hysteresis as the equilibrium FOT line potentially leads to an erroneous conclusion on thermodynamic quantities, such as field-induced latent heat accompanying the FOT, even at a qualitative level.

\subsection*{I\lowercase{ntroduction  to} (F\lowercase{e}$_{1-y}$Z\lowercase{n}$_{y}$)$_{2}$M\lowercase{o}$_{3}$O$_{8}$}
Our target material, (Fe$_{1-y}$Zn$_{y}$)$_{2}$Mo$_{3}$O$_{8}$, is a polar crystal (space group $P6_{3}mc$) with a linear magnetoelectric effect \cite{Kurumaji2015, Wang2015}. The magnetic properties are dominated by two kinds of Fe$^{2+}$ sites surrounded by oxygen tetrahedra (A-sites) and octahedra (B-sites) (Fig.~1{\bf c}); local spins of three Mo$^{4+}$ ions form a nonmagnetic spin-trimer singlet \cite{cotton1964} and thus have no contribution to the magnetism. According to a previous neutron study on the mother compound \FMO \cite{Bertrand1975}, the magnetic moments at octahedral sites are slightly larger than those at tetrahedral sites; hence, the spin configuration shown in Fig.~1{\bf d} has no net magnetization (antiferromagnetic; AFM), whereas that shown in Fig.~1{\bf e} creates an appreciable macroscopic magnetization (ferrimagnetic; FRI). The ground state at zero magnetic field is the AFM phase, which is replaced with the FRI phase when a sufficiently high magnetic field is applied. For undoped Fe$_{2}$Mo$_{3}$O$_{8}$, the magnetic-field-induced FOT between the AFM to the FRI phases occurs far above 15 T below 30 K \cite{Kurumaji2015}. 

The molecular-field analysis \cite{Bertrand1975} argues that the intralayer A--B, interlayer A--B and interlayer A--A magnetic interactions are all AFM, whereas the interlayer B--B magnetic interaction is weakly FM. In addition, the magnitudes of the interlayer A--B and interlayer A--A magnetic interactions are comparable, resulting in a delicate energy balance between the AFM and FRI phases in this system. Doped nonmagnetic Zn ions selectively occupy tetrahedral sites \cite{varret1972etude} and weaken the effective interlayer A--A magnetic interactions. Thus, the Zn-doping stabilizes the FRI phase, and accordingly the transition field to the FRI phase decreases upon Zn doping. In this study, we chose 5\% Zn-doping ($y = 0.05$) so that both the AFM and FRI phases are accessible in the feasible magnetic field range of 14 T \cite{Kurumaji2015}.

\subsection*{N\lowercase{otes on thermodynamic analysis}}
\subsubsection*{Clausius--Clapeyron equation}
On an equilibrium FOT line separating two phases, $\alpha$ and $\beta$, the thermodynamic potentials of the two phases are equal to each other by definition. This condition leads to the Clausius--Clapeyron equation, which is given in the following form for the case of an equilibrium FOT line in an $H$--$T$ phase diagram:
\begin{equation}
\frac{dT^{*}}{dH^{*}}=-\frac{M^{*}_{\rm \alpha}-M^{*}_{\rm \beta}}{S^{*}_{\rm \alpha}-S^{*}_{\rm \beta}}\equiv -\frac{\Delta M^{*}_{\rm \alpha\beta}}{\Delta S^{*}_{\rm \alpha\beta}},
\end{equation}
where ($H^{*}$, $T^{*}$) represents an arbitrary point on the equilibrium FOT line and $M^{*}_{\rm \alpha}$ (or $M^{*}_{\rm \beta}$) and $S^{*}_{\rm \alpha}$ (or $S^{*}_{\rm \beta}$) are the magnetization and entropy values of the $\alpha$ (or $\beta$) phase on the equilibrium FOT line, respectively. Note that any equilibrium FOT line in a $H$--$T$ phase diagram must satisfy this equation. This requirement means that by following the Clausius--Clapeyron equation, an equilibrium FOT line can be drawn sequentially by integrating $\frac{dT^{*}}{dH^{*}}$, as detailed in Results section.

To determine the slope of the FOT, $\frac{dT^{*}}{dH^{*}}$, the entropy and magnetization changes, $\Delta S^{*}_{\rm \alpha\beta}$ and $\Delta M^{*}_{\rm \alpha\beta}$, respectively, accompanying the magnetic-field-induced FOT should be given. To this end, the $S$--$H$ and $M$--$H$ curves should be derived for the $\alpha$ and $\beta$ phases, separately. Although the $M$--$H$ curves are readily measurable, the $S$--$H$ curves are not because the magnetic-field dependence of the specific heat, $c_{p}$, does not provide the $S$--$H$ curve. To obtain the $S$--$H$ curves and thus $\Delta S^{*}_{\rm \alpha\beta}$, we measure the $c_{p}(T)$ at predetermined magnetic fields: $H$ = 0, 0.1, 1, 2, 3, 4, 5, 6 and 7 T (we use the notation $H \equiv \mu_{0}H$  as an external magnetic field, where $\mu_{0}$ is magnetic permeability in vacuum). Thus, the $S_{H}(T)$ are derived by using $S_{H}(T)=\int_{0}^{T}\frac{c_{p}(H,T')}{T'}dT'$, and discretized $S_{T}(H)$ data points at a given $T$ are obtained.

\subsubsection*{Application of the Maxwell relation}
In principle, if the $S$--$T$ curves are given for many magnetic field points, isothermal $S$--$H$ curves at different temperatures can be derived with sufficient accuracy. However, from an experimental point of view, the measurement of specific heat requires much more time than that of magnetization; therefore, collecting $c_{p}(T)$ at many magnetic-field points is not practical. To overcome this technical difficulty, we refer to the Maxwell relation, $\left(\frac{\partial S}{\partial H}\right)_{T}=\left(\frac{\partial M}{\partial T}\right)_{H}$, which enables compensation for the limited number of $S_{T}(H)$ data points by referring to easily measurable magnetization data. Thus, isothermal $S$--$H$ curves are obtainable with improved accuracy.

For the purpose of deriving $\Delta S^{*}$ accompanying an equilibrium FOT, however, the Maxwell relation should be carefully used. To begin, the Maxwell relation cannot be used at an ideal FOT in principle because its derivation postulates that $S$ and $M$ are a differentiable continuous function of $H$ and $T$. This prerequisite is obviously not satisfied at an ideal FOT, which is accompanied by discontinuity in $S$ and $M$. One may assume that an actual FOT often exhibits relatively continuous change in $S$ and $M$ as a result of the evolution of the two-phase mixture (see Supplementary Fig.~1), and thus, the application of the Maxwell relation to the transition region appears fine. Nevertheless, the integration of $\left(\frac{\partial S}{\partial H}\right)_{T}=\left(\frac{\partial M}{\partial T}\right)_{H}$ with respect to $H$ across the transition region still tends to overestimate the intrinsic value of $\Delta S^{*}$; that is, the integration of the Maxwell relation across the FOT, frequently used in the analysis of the magnetocaloric effect\cite{franco2012,xu2015indirect}, cannot be applied to the precise determination of $\Delta S^{*}$ (for more details, see Supplementary Note 1). Therefore, we restrict the use of the Maxwell relation to when the magnetic state is considered a single phase of either the AFM or FRI phases. $\Delta S^{*}$ cannot be obtained from the Clausius-Clapeyron equation either \cite{xu2015indirect}, because the determination of the exact profile of the equilibrium FOT line itself is the purpose of this study.

\begin{figure}[htbp]
	\includegraphics[width=0.5\hsize]{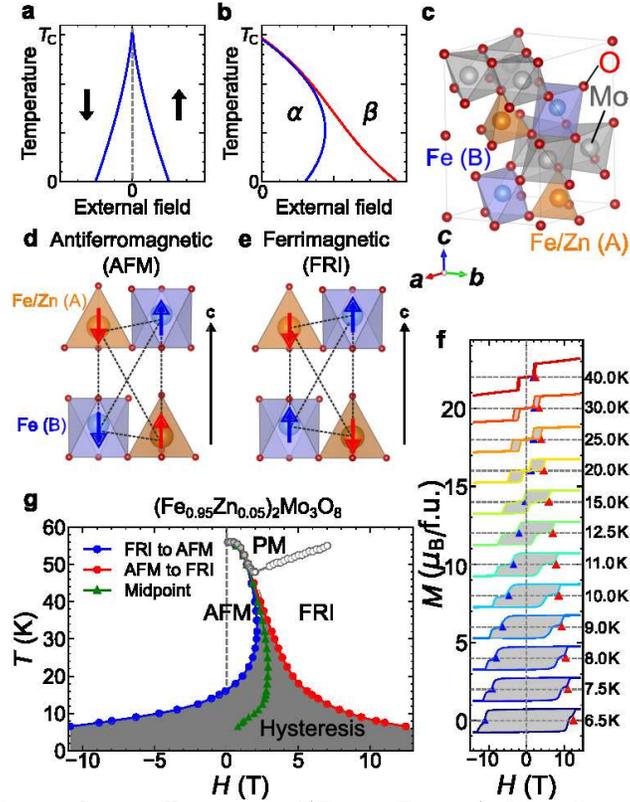}
	\vspace{-8mm}
	\caption{{\bf Field-temperature phase diagram of}$\mathbf{ (Fe_{0.95}Zn_{0.05})_{2}Mo_{3}O_{8}.}$ {\bf a,b} Phase diagrams with temperature and external field axes in the system with field-induced first-order phase transitions: ({\bf a}) ferroic-order case in which two states are connected with symmetry operations, such as time-reversal and space-inversion symmetries, and ({\bf b}) two-phase-competing case in which two thermodynamic phases indicated by $\alpha$ and $\beta$ are not degenerate in free energy at zero external field. The solid lines in (a) and (b) represent magnetic fields at which the phase transition is observed during magnetic field sweeps at a given rate. {\bf c} The crystal structures of (Fe$_{1-y}$Zn$_{y}$)$_{2}$Mo$_{3}$O$_{8}$. Fe$^{2+}$ ions at the A-sites (brown) and B-sites (blue) are surrounded by oxygen tetrahedra and octahedra, respectively. Mo$^{4+}$ ions (gray) form the nonmagnetic spin-trimer. {\bf d,e} The magnetic structure of the ({\bf d}) antiferromagnetic (AFM) and ({\bf e}) ferrimagnetic (FRI) phases. {\bf f} The isothermal magnetization curves of \FZMO at selected temperatures. Each curve is shifted by 2 $\mu_{\rm B}/{\rm f.u.}$ for clarity. The transition from the AFM to the FRI phases and from the FRI to the AFM phases are indicated by red and blue triangles, respectively. {\bf g} The temperature-magnetic-field phase diagrams of (Fe$_{0.95}$Zn$_{0.05}$)$_{2}$Mo$_{3}$O$_{8}$. The transitions from the AFM to the FRI phases and from the FRI to the AFM phases are indicated by red and blue circles, respectively. Gray closed and open circles represent the transition between PM and AFM and the crossover between PM to FRI, respectively, and they were determined from $M$--$T$ curves. Green triangles represent the midpoints between two transitions from the AFM (FRI) to the FRI (AFM) phases. In {\bf g,f}, the gray-hatched areas represents the hysteresis region.
	}
	\label{fig1}
\end{figure}

\section*{R\lowercase{esults}}
\subsection*{Field-temperature phase diagram}
Figure 1{\bf f} shows the isothermal magnetization curves of the target compound, (Fe$_{0.95}$Zn$_{0.05}$)$_{2}$Mo$_{3}$O$_{8}$. The red (blue) triangles indicate the transition fields from the AFM to the FRI phases (the FRI to the AFM phases) upon increasing (decreasing) magnetic field. The two hysteresis lines accompanying the magnetic-field-induced FOT in a positive magnetic field are drawn by tracing these two transition-fields. The experimental phase diagram is thus obtained, as shown in Fig.~1{\bf g}.

This material exhibits distinct hysteresis broadening at low temperatures. The hysteresis width, which is the difference between two transition fields, is as small as 0.4 T at 40 K; however, it significantly increases to $\approx$ 20 T at 6.5 K. In Fig.~1{\bf g}, the midpoints of the two hysteresis lines are plotted for reference. Note that the midpoint line starts to bend below 18 K, and its slope is appreciably positive at 6.5 K. If one considers the midpoint line as an approximate equilibrium FOT line, the midpoint line should bend again at low temperatures to be perpendicular to the magnetic-field axis to satisfy the third law of thermodynamics (Nernst--Planck hypothesis \cite{Callen1985}). Thus, the Clausius--Clapeyron equation concludes that \dSs ($\equiv S^{*}_{\rm FRI}-S^{*}_{\rm AFM}$) and/or \dMs ($\equiv M^{*}_{\rm FRI}-M^{*}_{\rm AFM}$) should exhibit complicated behaviour below 18 K. This finding, however, appears unusual, and its validity should be carefully considered by testing whether the midpoint line accurately represents the equilibrium FOT line.

\subsection*{$\bm{M}$--$\bm{T}$ curves}
We aimed to determine the equilibrium FOT line by following the equilibrium thermodynamics for a single phase. It was therefore important to avoid the data analysis in the $(H, T)$ region where the magnetic state is considered AFM--FRI phase mixture. In particular, extensive attention has to be paid to the magnetic state during cooling because we collected the $c_{p}(T)$ upon decreasing temperature. To identify the $(H, T)$ region of the two-phase mixture during cooling, we measured the $M$--$T$ curves at various magnetic fields during field-cooling (FC) and field-warming (FW) processes; in FC process, the magnetic fields were applied at 100 K ($> T_{\rm c} \approx$ 56 K) and the magnetization was measured from 100 to 2 K while retaining the magnetic fields; then, in FW process, the magnetization was measured from 2 to 100 K at the same magnetic fields. The data at selected magnetic fields are shown in Fig.~2{\bf a}. Overall, a transition from the PM to the AFM phase was observed below 1.5 T, whereas that to the FRI phase was observed above 3 T. At an intermediate field ranging from 1.6 to 3 T, a sharp, continuous change in magnetization and the associated thermal hysteresis were observed, signifying the thermally induced FOT between the AFM and FRI phases (see also Fig.~1{\bf g}).

\subsection*{Identification of the $\bm{(H, T)}$ region of single magnetic phase during field cooling}

\begin{figure}[htbp]
	\centering
	\vspace{-0mm}
	\includegraphics[width=0.5\hsize]{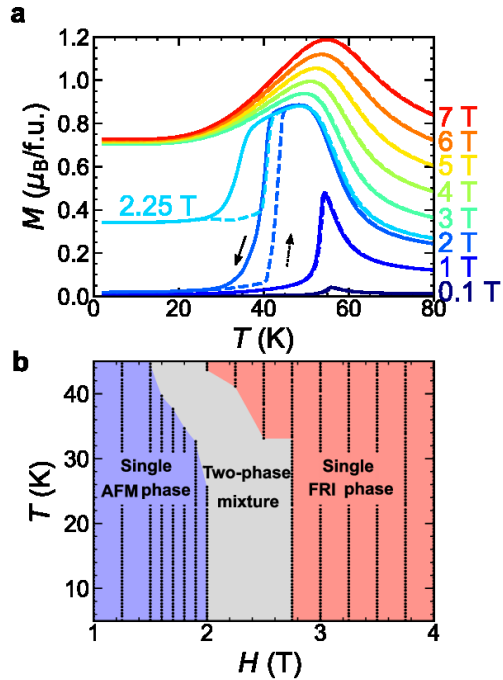}
	\vspace{0mm}
	\caption{{\bf Identification of single-phase AFM and FRI phases from $\bm{M}$-$\bm{T}$ curves.} {\bf a} Temperature dependencies of the magnetization upon field-cooling (FC; solid lines) and field-warming (FW; dashed lines) processes. {\bf b} Schematic phase diagram highlighting the single thermodynamic AFM and FRI phases. The black dots represent data points used for the present analysis while carefully avoiding the ``two-phase mixture'' (gray) region.}
	\label{fig2}
\end{figure}
\begin{figure}[htbp]
	\centering
	\vspace{-0mm}
	\includegraphics[width=0.5\hsize]{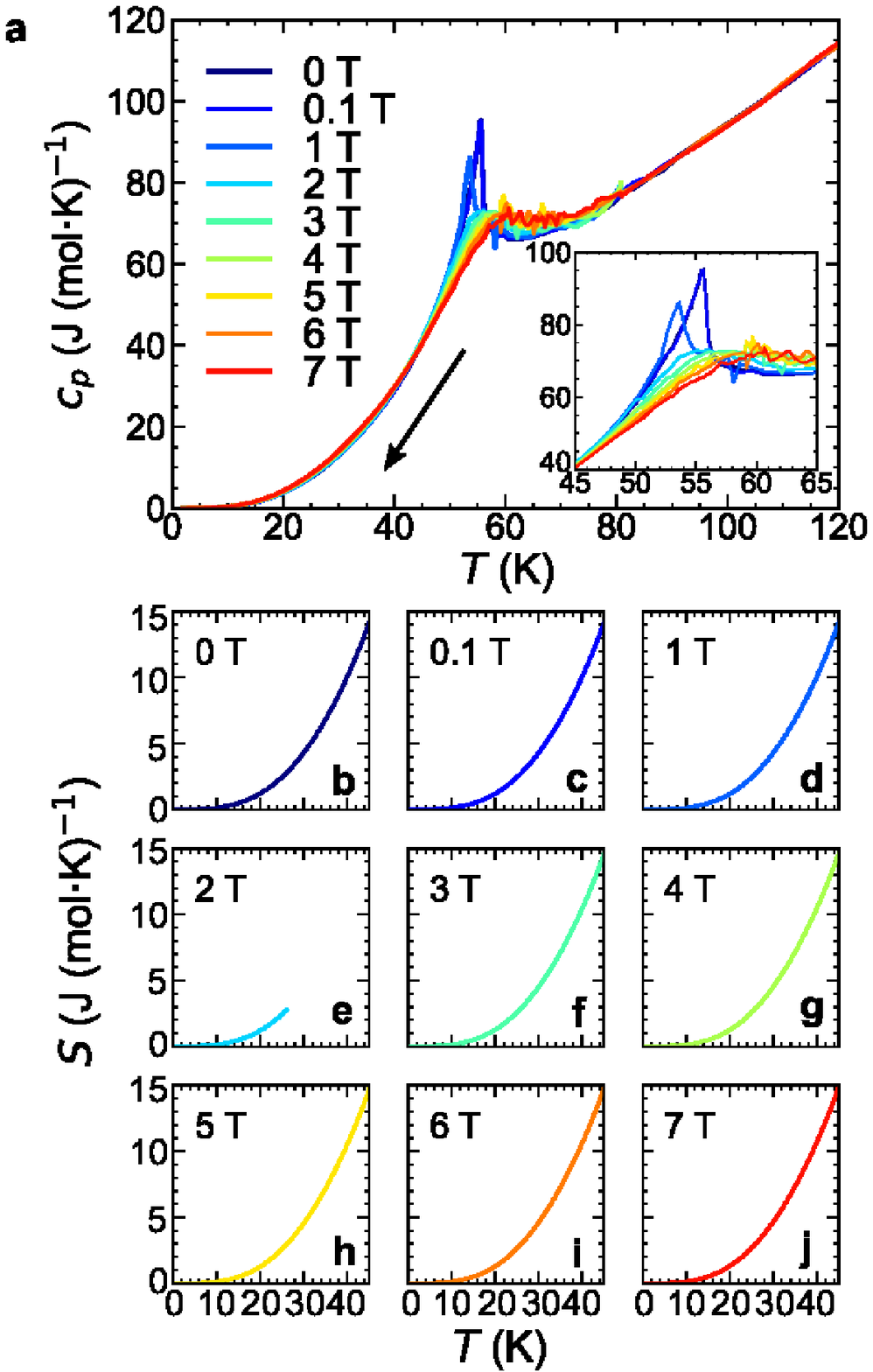}
	\vspace{0mm}
	\caption{{\bf Temperature dependences of the specific heats and entropies under magnetic fields.} {\bf a} Temperature dependencies of the specific heats under various magnetic fields. The specific heat measurement was performed during the field-cooling process. Inset: Enlarged view around the transitions. The $c_{p}(T)$ at 0 and 0.1 T are almost identical and no difference is distinguishable on this scale. {\bf b--j} The temperature dependencies of the entropies under various magnetic fields.}
	\label{fig3}
\end{figure}


From the $M$--$T$ curves, we determined the $(H, T)$ regions where the two-phase coexistence could occur during FC, as summarized in Fig.~2{\bf b}, according to the following two criteria. First, in the temperature range where the transition progressed, we determined that the data were affected by the two-phase coexistence. Second, especially below 30 K, even if the transition did not appear to progress, we determined that the two-phase mixture was present when its magnetization value exhibited an intermediate value between the AFM and FRI single phases. The lowest-temperature state below 2 T consisted exclusively of the AFM single phase (see Supplementary Note 2 and Supplementary Fig.~2, in which the $M$--$T$ curve under FW at 2 T after zero FC was compared with that under FC at 2 T). Similarly, above 3 T, the magnetization at the lowest temperature was insensitive to the magnetic field (see also Fig.~1{\bf f}), indicating the FRI single phase. In contrast, at 2.25 T, although thermal hysteresis was not observed below 25 K, the magnetization value at the lowest temperature was intermediate between the AFM and FRI single phase values. Thus, the magnetic state at 2.25 T below 25 K was considered the two-phase mixture.

The $(H, T)$ region that fell under ``two-phase mixture'' criteria is shown in grey in Fig.~2{\bf b}. The data from this region were not used in the following single-phase thermodynamic analysis to avoid a possible influence of the two-phase mixture.

\begin{figure*}[htbp]
	\centering
	\vspace{-0mm}
	\includegraphics[width=\hsize]{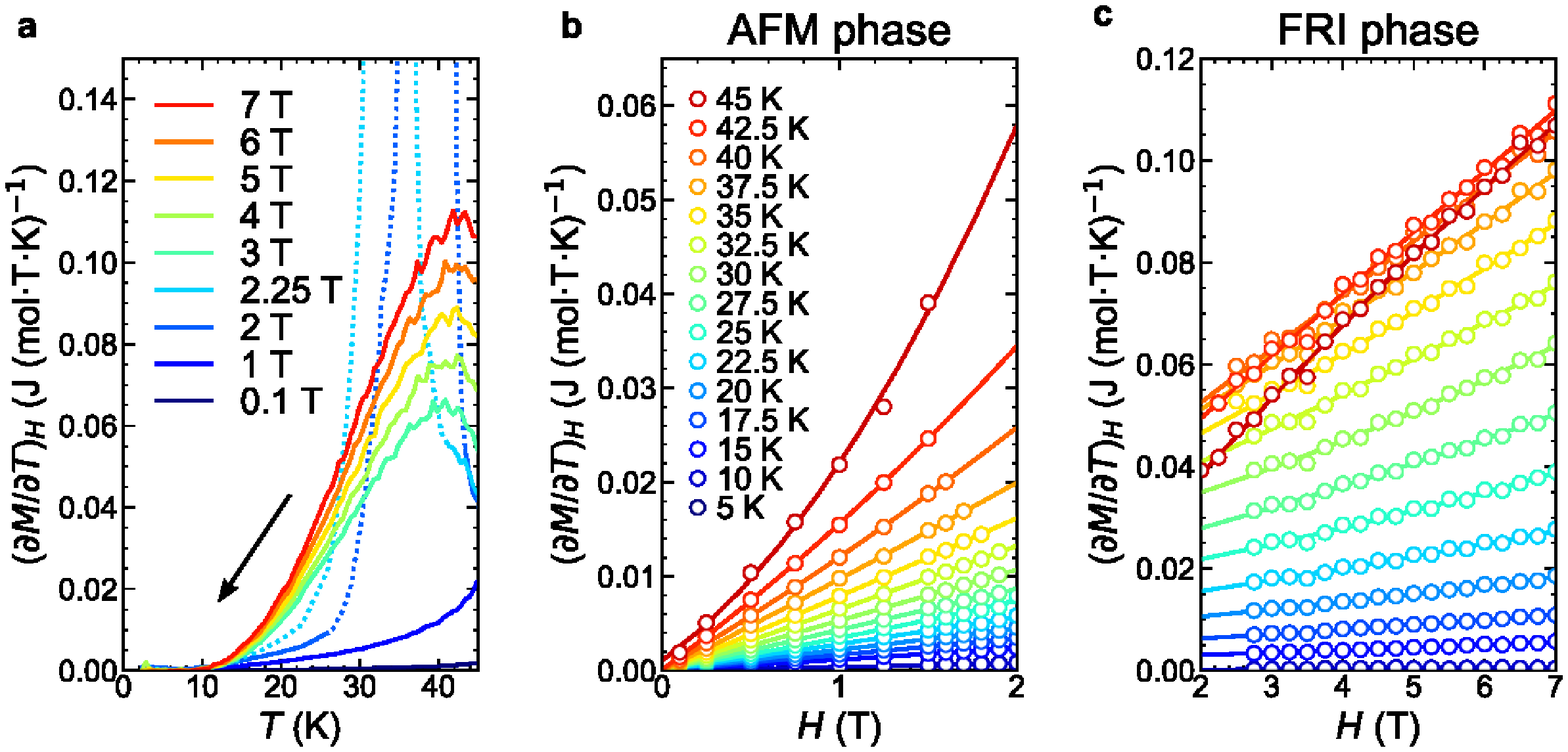}
	\vspace{0mm}
	\caption{{\bf Derivation of isothermal $\bm{\left(\frac{\partial S}{\partial H}\right)_{T}}$ curves via the Maxwell relation.} {\bf a} $\frac{dM}{dT}$--$T$ curves measured upon decreasing temperature. The data indicated by the dotted lines were not used for the Maxwell relation. {\bf b,c} Isothermal $\frac{dM}{dT}$--$H$ curves of the ({\bf b}) AFM and ({\bf c}) FRI phases. The solid lines represent the best-fit results (see the main text). The legend shown in ({\bf b}) also applies to ({\bf c}).}
	\label{fig4}
\end{figure*}

\begin{figure*}[htbp]
	\centering
	\vspace{-0mm}
	\includegraphics[width=\hsize]{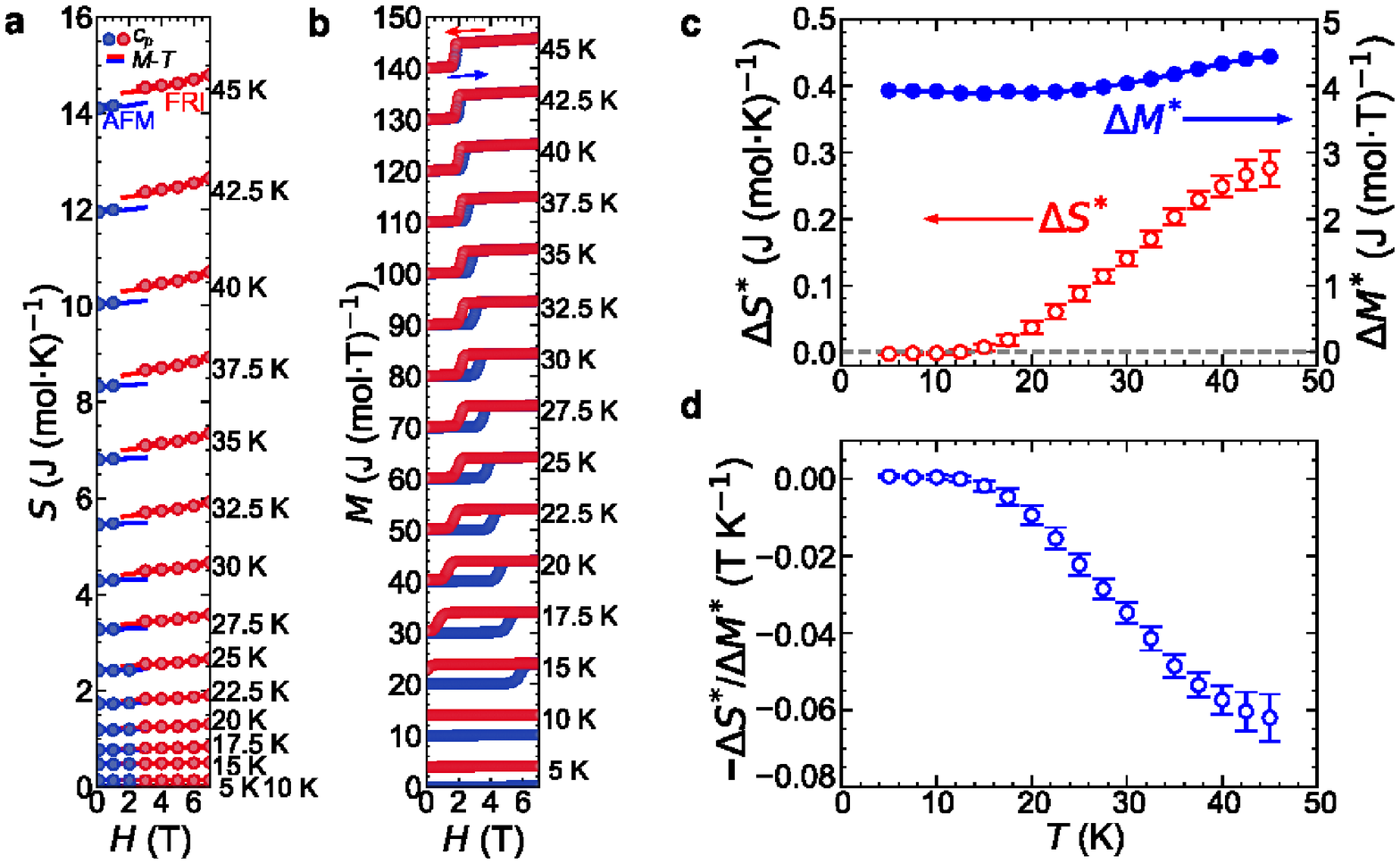}
	\vspace{0mm}
	\caption{{\bf Determination of $\bm{\Delta S^{*}}$ and $\bm{\Delta M^{*}}$.} {\bf a} Isothermal entropy curves at selected temperatures. Closed circles are the data obtained from the specific heat measurements. Blue and red solid lines indicate the fittings for the AFM and the FRI phases, respectively, based on the $M$--$T$ curve and the Maxwell relation. The error bars are estimated from the least-squares fitting. {\bf b} Isothermal magnetization curves at selected temperatures. Each magnetization curve is shifted by 10 J/(mol$\cdot$T) for visibility. {\bf c} Temperature dependence of the entropy and magnetization changes on the determined phase boundary. {\bf d} Temperature dependence of the ratio of \dSs and $\Delta M^{*}$. The error bars in ({\bf c}) and ({\bf d}) are estimated from the standard errors of the parameters when performing the least-squares fitting of the cubic function to the $\frac{dM}{dT}$ data.}
	\label{fig5}
\end{figure*}

\subsection*{Derivation of $\bm{S}$--$\bm{H}$ curves}
Figure 3{\bf a} shows the temperature dependences of the specific heat, $c_{p}$, at various magnetic fields measured upon field cooling. A clear lambda-like anomaly is observed in the $c_{p}(T)$ at approximately 55 K at 0 T, signifying a second-order transition from the paramagnetic (PM) to the AFM phases. The application of a magnetic field larger than 2 T suppresses the $\lambda$ peak, reflecting that a change from the PM to the FRI phases is crossover under a finite magnetic field (see Fig.~1{\bf g}). Similar behaviours have been reported in previous studies \cite{Wang2015, Ideue2017, Csizi2020}. The $S$--$T$ profiles are obtained by integrating $c_{p}/T$ with respect to $T$. As described in Supplementary Information (Supplementary Note 3), the value of entropy at 2 K is obtained at each magnetic field by extrapolating the $c_{p}(T)$ to zero temperature with $T^{3}$ behaviour. The resulting $S(T)$ data are shown in Figs.~3{\bf b--j}. Note that a discontinuous entropy change associated with an FOT cannot be accurately evaluated by integrating $c_{p}/T$ because $\frac{dS}{dT}=\frac{c_{p}}{T}$ is valid only when $dS/dT$ is well defined (i.e. $S$ is a differential continuous function of $T$). If the latent heat associated with the FOT can be accurately captured by specific heat measurements, the entropy above the transition temperature can be determined; however, this is not the case. Therefore, the $S$--$T$ curve (i.e., an entropy counting from zero temperature) at 2 T is only valid up to 27 K (Fig.~3{\bf e}). From these $S_{H}(T)$ data set, the discretized $S_{T}(H)$ points are derived (for instance, see Fig.~5{\bf a}). The comparison of $S_{H}(T)$ between 0 and 7 T tells that the FRI phase has a higher entropy than the AFM phase (Supplementary Fig.~4).

Then, we aim to derive the $S$--$H$ curves with accuracy from the limited number of available $S_{T}(H)$ data points by using the Maxwell relation. We derive the $\left(\frac{\partial M}{\partial T}\right)_{H}-T$ curves from the $M$--$T$ curves arising from a single-phase region (either the AFM or FRI phases). The results at selected magnetic fields are shown in Fig.~4{\bf a} with solid lines (for the data above 50 K, see Supplementary Note 4). Then, from the continuous $\left(\frac{\partial M}{\partial T}\right)_{H}-T$ curves, we obtain the discretized $\left(\frac{\partial M}{\partial T}\right)_{H}$ data points as functions of $H$ at selected temperatures for the AFM and FRI single phases, as shown in Figs.~4{\bf b} and 4{\bf c}, respectively. 

To obtain the isothermal $S$--$H$ curves, $\left(\frac{\partial S}{\partial H}\right)_{T}$ should be integrated with respect to $H$. To this end, we first fit the magnetic field dependence of $\left(\frac{\partial M}{\partial T}\right)_{H}=\left(\frac{\partial S}{\partial H}\right)_{T}$ for the AFM and FRI phases separately by the following polynomial functions:

\begin{widetext}
	\begin{equation}
	\left(\frac{\partial M_{\rm AFM}(H,T)}{\partial T}\right)_{H}=
	\left(\frac{\partial S_{\rm AFM}(H,T)}{\partial H}\right)_{T}=
	a_{3}(T)H^{2}+a_{2}(T)H+a_{1}(T)
	\end{equation}
	\begin{equation}
	\left(\frac{\partial M_{\rm FRI}(H,T)}{\partial T}\right)_{H}=
	\left(\frac{\partial S_{\rm FRI}(H,T)}{\partial H}\right)_{T}=
	b_{3}(T)(H-7)^{2}+b_{2}(T)(H-7)+b_{1}(T).
	\end{equation}
\end{widetext}

\noindent The fittings are successful as shown in Figs.~4{\bf b} and 4{\bf c}, from which the parameters $a_{i}$ and $b_{i}$ ($i = 1, 2, 3$) are determined. The $S$--$H$ curves are thus given as follows:

\begin{widetext}
	\begin{equation}
	S_{\rm AFM}(H,T) = S_{\rm AFM}(0,T) 
	+ \int_{0}^{H} \left(\frac{\partial S_{\rm AFM}}{\partial H'} \right)_{T}dH'=
	\frac{a_{3}(T)}{3}H^{3}+\frac{a_{2}(T)}{2}H^{2}+a_{1}(T)\mu_{0}H+a_{0}(T),
	\label{sfit1}
	\end{equation}
	\begin{equation}
	S_{\rm FRI}(H,T) = S_{\rm FRI}(7,T) 
	+ \int_{7}^{H} \left(\frac{\partial S_{\rm FRI}}{\partial H'} \right)_{T}dH'=
	\frac{b_{3}(T)}{3}(H-7)^{3}+\frac{b_{2}(T)}{2}(H-7)^{2}+b_{1}(T)(H-7)+b_{0}(T).
	\label{sfit2}
	\end{equation}
\end{widetext}

\noindent Note that $a_{i}$ and $b_{i}$($i$=1,2,3) are given from the fitting of the $\left(\frac{\partial M}{\partial T}\right)_{H}$ data points. Finally, to determine the remaining parameters $a_{0}$ (for the AFM phase) and $b_{0}$ (for the FRI phase), we refer to the discretized $S_{T}(H)$ points, which are independently obtained from the $c_{p}$--$T$ curves. Because there is only one unknown parameter for each phase, the problem of the limited $S_{T}(H)$ points is minimized. As shown in Fig.~5{\bf a}, the fitting $S$--$H$ curves based on the magnetization data, equations~(\ref{sfit1}) and (\ref{sfit2}), well reproduce the $S_{T}(H)$ data points, which are obtained from the specific heat measurement. This good agreement corroborates that the present method based on the single-phase thermodynamics successfully avoids the experimental artifacts in the derived $S$--$H$ curve\cite{liu2007determination,tocado2009entropy,amaral2010estimating} that are often caused by a hysteretic behaviour and a kinetic effect associated with a FOT  (see Supplementary Note 1). In this way, the continuous $S$--$H$ curves are derived at various temperatures for the AFM and FRI single phases, separately. The resulting two curves are not connected to each other, representing the entropy discontinuity accompanying the FOT. Overall, the entropy difference decreases as the temperature decreases. Nevertheless, to determine $\Delta S^{*}$ more precisely, the equilibrium transition field, $H^{*}$, should be determined at each temperature. This issue is discussed below.

\subsection*{$\bm{M}$--$\bm{H}$ curves}
To apply the Clausius–Clapeyron equation, $\Delta M^{*}$ should also be derived from the isothermal $M$--$H$ curves. In contrast to the $S$--$H$ curves, the $M$--$H$ curves are straightforwardly measured, as shown in Fig.~5{\bf b}. To obtain the magnetization of a single phase, we first prepare the AFM and FRI single phases by ZFC and 7 T-FC from 100 K ($> T_{\rm c} \approx 56$ K), respectively; then, the magnetization is measured while increasing and decreasing the magnetic fields, respectively;  the cooling process from 100 K is performed after each $M$--$H$ measurement. The two branches are separated from each other, and the transition to the other branch occurs when the magnetic field reaches a value of the hysteresis line (Figs.~1{\bf f,g}). From these $M$--$H$ curves, one can determine $\Delta M^{*}$, if the equilibrium transition field $H^{*}$ is given at each temperature.

\subsection*{Derivation of the equilibrium first-order transition line}
The determination of \dSs and \dMs and that of the equilibrium transition field \Hs are an intertwined issue. If \Hs is given at a certain temperature, \dSs and \dMs at the same temperature are determined; if \dSs and \dMs are given at a certain temperature, \Hs at a nearby temperature is determined. To perform this sequential determination, the starting point for drawing the equilibrium FOT line must be determined first.

The starting point is chosen at ($H^{*}_{0}$, $T^{*}_{0}$) = (1.81 T, 45 K) because the hysteresis width is as small as 0.1 T and it is reasonable to approximate the midpoint of the two hysteresis lines as the equilibrium FOT point. First, the values of \dSs and \dMs at (1.81 T, 45 K) are determined by referring to the isothermal $S$--$H$ and $M$--$H$ curves, respectively; the slope of the FOT line at ($H^{*}_{0}$, $T^{*}_{0}$), $\frac{dH^{*}}{dT^{*}}|_{T^{*}_{0}}$, is calculated from $-\frac{\Delta S^{*}}{\Delta M^{*}}$. The next transition field $H^{*}_{1}$ at the nearby temperature $T^{*}_{1}$ ($< T^{*}_{0}$) is calculated as follows:
\begin{align}
H^{*}_{1} &= H^{*}_{0} + \left(\frac{dH^{*}}{dT^{*}}\right)_{T^{*}_{0}}\times (T^{*}_{1}-T^{*}_{0}) \nonumber \\
&= H^{*}_{0} - \left(\frac{\Delta S^{*}}{\Delta M^{*}}\right)_{T^{*}_{0}, H^{*}_{0}}\times (T^{*}_{1}-T^{*}_{0})
\end{align}
We repeat this procedure sequentially, and the transition field $H^{*}_{n}$ at $T^{*}_{n}$  ($n \geq 1$) is obtained by the following equations:
\begin{align}
\Delta S^{*}(H^{*}_{n},T^{*}_{n}) &= S_{\rm FRI}(H^{*}_{n},T^{*}_{n}) - S_{\rm AFM}(H^{*}_{n},T^{*}_{n})\\
\Delta M^{*}(H^{*}_{n},T^{*}_{n}) &= M_{\rm FRI}(H^{*}_{n},T^{*}_{n}) - M_{\rm AFM}(H^{*}_{n},T^{*}_{n})\\
H^{*}_{n+1} &= H^{*}_{n} + \left(\frac{dH^{*}}{dT^{*}}\right)_{T^{*}_{n}}\times (T^{*}_{n+1}-T^{*}_{n}) \nonumber \\
&= H^{*}_{n} - \left(\frac{\Delta S^{*}}{\Delta M^{*}}\right)_{T^{*}_{n}, H^{*}_{n}}\times (T^{*}_{n+1}-T^{*}_{n})
\end{align}
The decrement of the temperature step, $T^{*}_{n+1}-T^{*}_{n}$, is determined by the temperatures at which the isothermal $M$--$H$ curves are measured, and it is $-$2.5 K in the present study. Thus, \dSs and \dMs, $-\frac{\Delta S^{*}}{\Delta M^{*}}$, and the equilibrium FOT line are determined sequentially for temperatures below 45 K, as displayed in Figs.~5{\bf c,d} and Fig.~6, respectively. \dSs is sensitive to temperature and monotonically decreases to zero as the temperature decreases to zero temperature, whereas \dMs depends only weakly on temperature (Fig.~5{\bf c}). Thus, the ratio, $-\frac{\Delta S^{*}}{\Delta M^{*}}$, monotonically approaches to zero at zero temperature (Fig.~5{\bf d}).

The obtained equilibrium FOT line (Fig.~6) becomes appreciably distinct from the midpoints of the two hysteresis lines, especially below 18 K. Note that the hysteresis distinctly broadens below 18 K. Toward zero temperature, the obtained equilibrium FOT line monotonically becomes perpendicular to the magnetic-field axis to satisfy the third law of thermodynamics. The time evolution measurements of the net magnetization, $M(t)$, indicates that the equilibrium phase at 2.3 T and 10 K is the AFM phase, rather than the FRI phase (Supplementary Note 5 and Supplementary Fig.~6), further confirming that the profile of the obtained equilibrium FOT line is consistent with the observation. The asymptotic behaviour of the FOT line indicates that nothing peculiar is involved below 18 K, as demonstrated in Figs.~5{\bf c} and 5{\bf d}, in contrast to the implications drawn from the profile of the midpoint line. 

\begin{figure}[htbp]
	\centering
	\vspace{-0mm}
	\includegraphics[width=0.5\hsize]{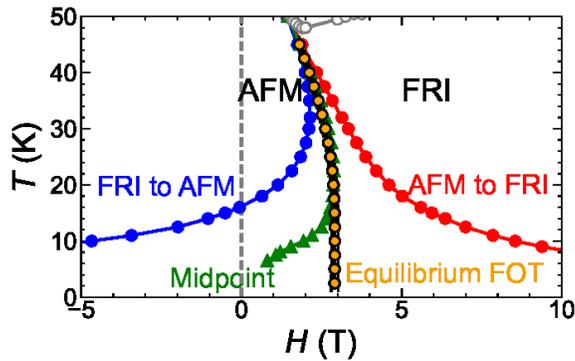}
	\vspace{0mm}
	\caption{{\bf Determination of the equilibrium first-order transition line.} Determined phase boundary indicated by orange circles with error bars, which are superimposed on the low-temperature enlarged view of Fig.~1{\bf g}.}
	\label{fig6}
\end{figure}

\section*{D\lowercase{iscussion}}

\begin{figure}[htbp]
	\centering
	\vspace{-0mm}
	\includegraphics[width=0.5\hsize]{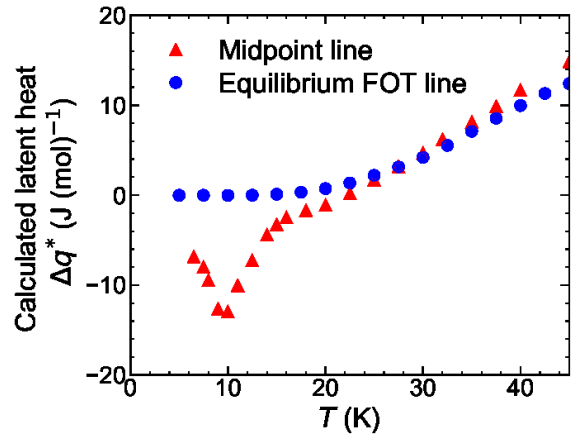}
	\vspace{0mm}
	\caption{{\bf The comparison of calculated latent heats between midpoint and equilibrium first-order-transition lines.} Temperature dependences of the calculated latent heats based on the midpoint line (red triangles) and the equilibrium FOT line (blue circles).}
	\label{fig7}
\end{figure}

	Finally, we show that if the midpoint line is regarded as the equilibrium FOT line, it can lead to even qualitatively incorrect conclusion regarding thermodynamic quantities such as latent heat. By using equation~(1), the latent heat $\Delta q^{*}$ accompanying the FOT from the AFM to the FRI can be calculated as $\Delta q^{*}=T^{*}\Delta S^{*}=-T^{*}\Delta M^{*}\left(\frac{dH^{*}}{dT^{*}}\right)$ for either the midpoint or equilibrium FOT lines. The calculated values of the latent heat for each case are compared in Fig.~\ref{fig7}. At high temperatures above 25 K, $\Delta q^{*}$ is positive, and its values derived from the midpoint and equilibrium FOT lines agree well with each other. However, the two curves show distinct behaviour below 25 K: $\Delta q^{*}$ derived from the equilibrium FOT line approaches zero asymptotically toward zero temperature, whereas that derived from the midpoint line shows nonmonotonous temperature dependence and even a large negative value, $\approx -$14 J/mol at 10 K. Thus, the adoption of the midpoint line as the equilibrium FOT line causes the erroneous latent heat even at a qualitative level, especially at low temperatures, demonstrating that caution should be exercised when referring to the midpoint lines. 

	The experimental observation of FOTs is known to be susceptible to kinetic effects, which are beyond the framework of the equilibrium thermodynamics. We have demonstrated that the equilibrium FOT line can nevertheless be determined using the single-phase thermodynamics with avoiding analysis of the phase-mixed state. The understanding of the temperature-dependent agreement/disagreement between the midpoint and equilibrium FOT lines remains an open question in this study. The large deviation of the midpoint line from the equilibrium line at low temperatures appears to be related to distinct hysteresis broadening. This observation may suggest that the details of the phase evolution in the FOT may vary with temperature. 

\clearpage
\section*{D\lowercase{ata availability}}
The datasets used and/or analysed during the current study available from the corresponding author on reasonable request.

\section*{M\lowercase{ethods}}
\subsection*{Sample preparation}A single crystal of \FZMO was grown by the chemical-vaper transport method. The details about the growth of this sample were reported in the literature \cite{strobel1982growth,strobel1983growth,Kurumaji2015}. All magnetization and specific heat measurements were performed using the same sample with a relatively small mass ($\approx$ 3 mg).

\subsection*{Magnetization measurement}
The magnetization along the $c$--axis was measured in Quantum Design MPMS-XL, MPMS-3, and PPMS 14 T.

\subsection*{Specific heat measurement}Specific heat measurements were performed in a field-cooling procedure, and the standard relaxation method available in the heat capacity option of Quantum Design PPMS 14 T was used. A small amount of grease (Apiezon N) was used to ensure good thermal contact between the sample stage (platform) and the sample. 

\clearpage

\bibliographystyle{./naturemag_revised}
\bibliography{./ref_short_italic}

\section*{\uppercase{a}\lowercase{cknowledgments}}
	The authors acknowledge enlightening discussions with Y.~Taguchi and S.~Imajo. This work was supported by JSPS KAKENHI (Grant No.~21K14398, No.~21H04442, and No.~18H05225) and JST CREST (No.~JPMJCR1874). K.M. was supported by the Special Postdoctoral Researcher Program of RIKEN. The crystal structure was visualized by VESTA 3 \cite{momma2011vesta}.

\section*{\uppercase{a}\lowercase{uthor contributions}}
K.M. and Y.N. performed the magnetization measurement and analyzed the data. K.M. and Y.N. performed the specific heat measurement and analyzed the data, with the help of M.K. T.K. grew the single crystal used for this study. K.M. and F.K. wrote the manuscript. All authors have discussed the results and commented on the manuscript.

\section*{\uppercase{c}\lowercase{ompeting interests}}
The authors declare no competing interests.





\end{document}